\documentclass[prd,nofootinbib,preprint,epsfig]{revtex4}
\usepackage{psfig}
\usepackage{amssymb}
\def\be{\begin{equation}}
\def\ee{\end{equation}}
\def\bea{\begin{eqnarray}}
\def\beas{\begin{eqnarray*}}
\def\eea{\end{eqnarray}}
\def\eeas{\end{eqnarray*}}
\def\ba{\begin{array}}
\def\ea{\end{array}}
\def\lsim{\mathrel{\rlap{\lower4pt\hbox{\hskip1pt$\sim$}}
    \raise1pt\hbox{$<$}}}      %less than or approx. symbol
\def\gsim{\mathrel{\rlap{\lower4pt\hbox{\hskip1pt$\sim$}}
    \raise1pt\hbox{$>$}}}      %greater than or approx. symbol
\begin{document}
%\preprint{\hbox{YITP-SB-54-03}}

\title{Testing Non--commutative QED $\gamma\gamma\gamma$ and 
$\gamma\gamma\gamma\gamma$ Couplings at LHC}

\author{S.~M.~Lietti \footnote{E-mail: lietti@fma.if.usp.br}}
\affiliation{Instituto de F\'{\i}sica da USP, C.P. 66.318, S\~ao Paulo, 
SP 05389-970, Brazil.}

\author{C.~A.~de~S.~Pires \footnote{E-mail: cpires@fisica.ufpb.br}}
\affiliation{Departamento de F\'{\i}sica da UFPb, C.P. 5008, Jo\~ao Pessoa, 
PB 58051-970, Brazil.}

%---------------------------------------------------------------------

\begin{abstract}
\vspace*{0.5cm}
In this work, we investigate the sensitivity of the process  
$p + p \to q + q \to j + j + \gamma + \gamma  $ at LHC for the 
photonic 3- and 4- point functions that appear in non--commutative QED. 
We show that this process serves to study the behavior of the 
space-space as well as of the space-time non--commutativity. 
We also show that this process can probe the 
non--commutative scale $\Lambda$ in the range of few TeV´s.
\end{abstract}

\maketitle

%---------------------------------------------------------------------

\section{Introduction}
\label{sec1}
The central idea behind non--commutative space-time (NCST) is that there must 
be a regime of energy where space-time loses its condition of continuum 
and passes to obey the relation $[\hat{x}_\mu ,\hat{ x}_\nu ] = 
i\frac{C_{\mu \nu}}{\Lambda^2}$~\cite{snyder}, where $C_{\mu \nu}$ is a 
real antisymmetric constant matrix. In other words, there must be a very 
microscopic region of space-time, or very high energy, where our common 
understanding of space-time is not applicable anymore. 

When people began to develop such idea, the scale of energy $\Lambda$ 
where non--commutativity was expected to manifest was around the Planck 
scale $M_{P\ell}$~\cite{connes,s-w}. This scale of energy is quite out of 
the present phenomenological reach. However, inspired by this recent 
idea of extra dimensions~\cite{extradim}, which suggest the fundamental 
Planck scale can be around few TeV´s, people  brought $\Lambda$ down to 
TeV scale~\cite{rizzo}. This leaves the idea of NCST phenomenologically 
attainable. In this regard, it turn out to be interesting reformulate the 
phenomenological models in the basis of the NCST. 

Unfortunately, the implementation of NCST is still a challenge for model 
building and presently the only consensual phenomenological model is 
non--commutative QED (NCQED)~\cite{rizzo,QEDphen,mathews,review}. 
The phenomenology of NCQED has been intensively investigated
~\cite{rizzo, QEDphen,mathews,review}. What has particularly 
called the people attention in NCQED is the photonic 3- and 4- point 
functions. The processes where such couplings appears are the Compton 
scattering, pair annihilation process, $e^+ e^- \rightarrow \gamma
\gamma$, and the $\gamma \gamma \rightarrow \gamma \gamma$. 

However, none process involving quarks was investigate yet. 
The reason is that in NCQED, the covariant derivatives can only be 
constructed for fermionic fields of charges $0$ and $\pm 1$. Therefore,  
the non--commutative photon-quark-quark interaction cannot be described by
the model. In order to solve this problem, people began to implement
the NCST effects into the Standard Model of particles. Two proposals for 
Non--commutative Standard Model (NCSM) can be found in the literature, 
one is based on the $U_{\star}(3)\times U_{\star}(2)\times U_{\star}(1)$ 
gauge group~\cite{ncsm1} while the other is based on the standard 
gauge group $SU(3)_C \times SU(2)_L \times U(1)_Y$, making use 
of the Seiberg-Witten maps~\cite{ncsm2}. Some phenomenology 
of these models are found in Refs.~\cite{ncsmphen}. However, 
no agreement has been reached yet regarding a phenomenological NCSM. 

Therefore, our main goal in this work is to study the potential of the 
LHC to only probe NCQED, particularly the  photonic 3- and 4- point 
functions $\gamma \gamma \gamma$ and $\gamma \gamma \gamma \gamma$, 
through the process $p + p \to q + q \to j + j + \gamma + \gamma$. 
In order to do so, the only assumption we took is that possible 
non--commutative quark-quark-photon interaction generates negligible
effects, allowing us to consider only the the standard model 
quark-quark-photon interaction and NCQED in our analysis. 

This work is organized as follow. In Sec.~(\ref{sec2}) we present the photonic 
3- and 4- point functions in NCQED. After in Sec.~(\ref{sec3}) we 
discuss the NCQED signal and the SM background. 
In Sec.~(\ref{sec4}), we finish with our conclusions.

\section{Photonic 3- and 4- point functions in NCQED}
\label{sec2}
One manner of settling  non--commutative coordinates in the  context 
of field theory is through the Moyal product, or the $\star$ product, 
whose expansion is~\cite{connes}  
\begin{eqnarray}
A(x) \star  B(x) \equiv [ e^{(i/2)\theta_{\mu \nu}\partial_{\zeta\mu} 
\partial_{\eta \nu}}A(x+\zeta)B(x+\eta)]_{\zeta=\eta=0}.
\label{operatorproduct}
\end{eqnarray}
With this product, we  procedure in the following way. We first 
formulate the  Lagrangian in  terms of  $\star$ product and then 
change the   $\star$ product by the expansion in (\ref{operatorproduct}) 
in order to leave the Lagrangian in terms of  ordinary product. 

In gauge theories first thing to do is to express the gauge transformation 
in terms of $\star$ products
\begin{eqnarray}
A_\mu \rightarrow U\star A_\mu  \star U^{-1} + 
\frac{i}{g}U\star \partial_\mu U^{-1}.
\label{gaugeinv1}
\end{eqnarray}
In the particular case of NCQED, where 
$U(x)=exp \star (ig\alpha(x))$, we have
\begin{eqnarray}
A_\mu \rightarrow A_\mu +\partial_\mu \alpha-
ig(A_\mu \star \alpha -\alpha \star A_\mu ).
\label{giqed}
\end{eqnarray}
 
In order to the action of the  NCQED preserves the gauge invariance, 
the tensor $F^*_{ \mu \nu}$ must present the form
\begin{eqnarray}
F^*_{\mu \nu}=\partial_\mu A_\nu -\partial_\nu A_\mu
-ig[A_\mu,A_\nu]_\star=F_{\mu \nu}
-ig( A_\mu \star A_\nu -A_\nu \star A_\mu).
\label{tensor}
\end{eqnarray}
With these expansions, the photonic part of the NCQED presents the 
following Lagrangian
\begin{eqnarray}
{\cal L}=&& -\frac{1}{4}F^{\mu \nu}F_{\mu \nu} 
- 2e\sin(\frac{p_1 C p_2}{2\Lambda^2})
(\partial_\mu A_\nu -\partial_\nu A_\mu)A^\mu A^\nu\nonumber \\
&&-4e^2\sin^2(\frac{p_1 C p_2}{2\Lambda^2})A^4,
\label{action}
\end{eqnarray}
 where $p_1 C p_2= p_1^\mu p_2^\nu C_{\mu \nu} $ and with 
$F_{\mu \nu}=\partial_\mu A_\nu -\partial_\nu A_\mu$ being the 
standard electromagnetic tensor. 
The Feynman rules for the vertices
$\gamma \gamma \gamma$ and $\gamma \gamma \gamma \gamma$ 
are given by~\cite{rizzo}, 
\begin{eqnarray}
\gamma_\mu(p_1) \gamma_\nu(p_2) \gamma_\rho(p_3)  \; &:& \;
2 g \sin(\frac{p_1Cp_2}{2 \Lambda^2}) [(p_1-p_2)^\rho g^{\mu \nu} +
(p_2-p_3)^\mu g^{\nu \rho} + (p_3-p_1)^\nu g^{\mu \rho} ] \; ;
\nonumber \\
\gamma_\mu(p_1) \gamma_\nu(p_2) \gamma_\rho(p_3) \gamma_\sigma(p_4) \; &:& \;
4 i g^2 [
 (g^{\mu\sigma} g^{\nu\rho} - g^{\mu\rho} g^{\nu\sigma} )
 \sin(\frac{p_1Cp_2}{2 \Lambda^2})\sin(\frac{p_3Cp_4}{2 \Lambda^2}) 
\nonumber \\
&&+(g^{\mu\rho} g^{\nu\sigma} - g^{\mu\nu} g^{\rho\sigma} ) 
\sin(\frac{p_3Cp_1}{2 \Lambda^2})\sin(\frac{p_2Cp_4}{2 \Lambda^2}) 
\label{feynman_rules} \\
&&+g^{\mu\nu} g^{\rho\sigma} - g^{\mu\sigma} g^{\nu\rho} () 
\sin(\frac{p_1Cp_4}{2 \Lambda^2})\sin(\frac{p_2Cp_3}{2 \Lambda^2}) ] \; ,
\nonumber 
\end{eqnarray}
where all the momenta are out-going.

The parametrization suggested by Hewett-Petriello-Rizzo~\cite{rizzo} 
for the antisymmetric matrix $C$  is
 \begin{eqnarray}
        C_{\mu \nu}=\left(\begin{array}{cccc} 
 0 & \sin \alpha \cos \beta &  \sin \alpha \sin \beta & \cos \alpha\\
 -\sin \alpha \cos \beta & 0 & \cos \gamma & -\sin \gamma  \sin \beta \\
 - \sin \alpha \sin \beta  & -\cos \gamma & 0 & -\sin \gamma \cos \beta \\
  -\cos \alpha & \sin \gamma \sin \beta & \sin \gamma \cos \beta & 0
\end{array}
\right),
\label{matrixC}
\end{eqnarray}
where the three  angles used to parametrize  $C_{\mu \nu}$ are related with 
the direction of the background {\bf E} and {\bf B}-fields. In this 
parametrization, the angle $\beta$ define the origin of the $\phi$ 
axis~\cite{rizzo}. The common procedure here is to fix $\phi$ by 
settling $\beta =\pi/2$. Therefore, the antisymmetric matrix 
get parametrized by two angles: the angle $\alpha$ related 
to the space-time non--commutativity, and the angle $\gamma$  
related to the space-space non--commutativity. 

\begin{figure}
\protect
\centerline{\mbox{\psfig{file=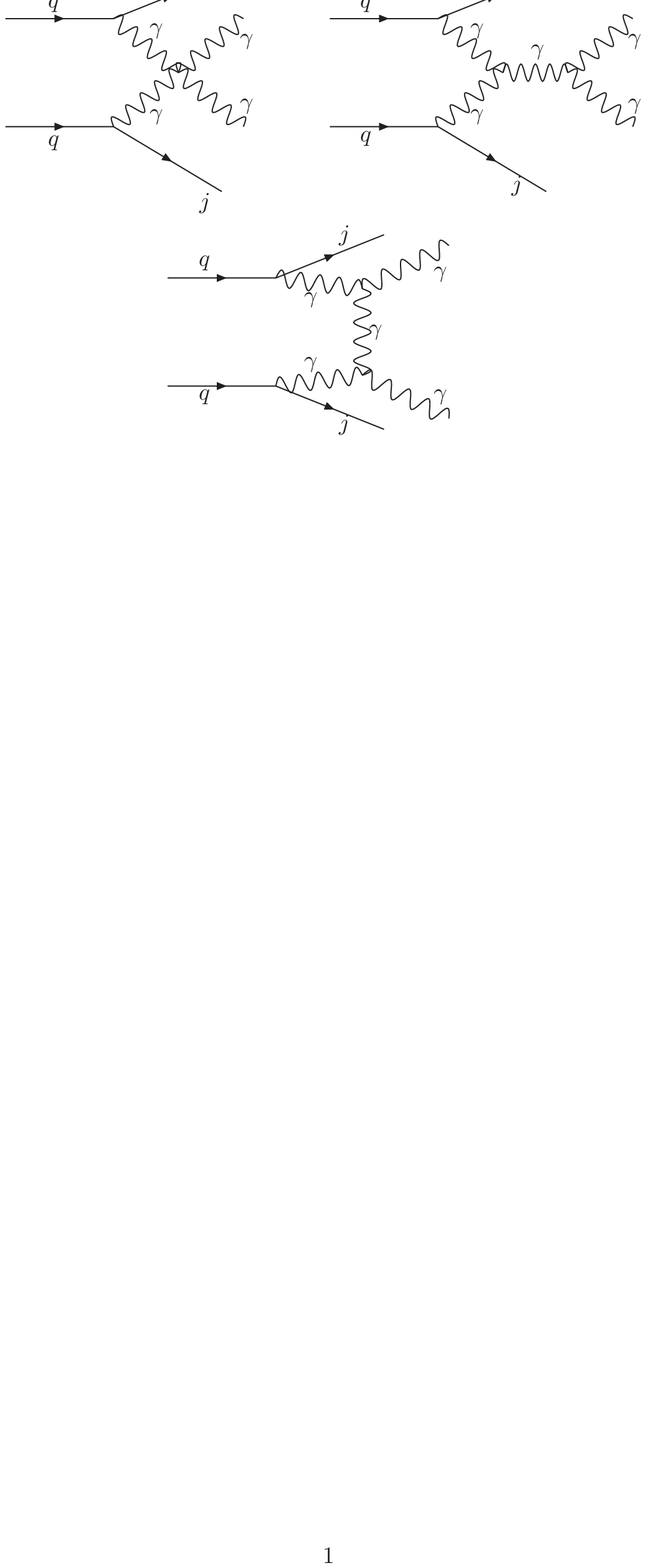,angle=0,width=1.\textwidth,clip=}}}
\vskip -1. cm 
\caption{NCQED photonic 3- and 4- point functions contributions 
for the reaction~(\ref{jj}).}
\label{fig01}
\end{figure}

In order to test the NCQED vertexes given by equations~(\ref{feynman_rules}), 
we perform a detailed analysis of the production via weak boson
fusion(WBF) of photon pairs accompanied by jets, {\em  i.e.},
\begin{equation}
p + p \to q + q \to j + j + \gamma + \gamma \; .
\label{jj}
\end{equation}
Beyond the expected SM Feynman diagrams, reaction~(\ref{jj}) receives 
contributions from NCQED photonic 3- and 4- point functions, shown in 
Fig.~\ref{fig01}.

The advantage of WBF, where the scattered final-state quarks receive 
significant transverse momentum and are observed in the detector as 
far-forward/backward jets, is the strong reduction of QCD backgrounds 
due to the kinematical configuration of the colored part of the event.

%---------------------------------------------------------------------
\section{Signals and backgrounds}
\label{sec3}

In this section we study the reaction (\ref{jj}) at the LHC. We evaluated 
numerically the helicity amplitudes of all the SM subprocesses leading to 
the $jj \gamma\gamma$ final state where $j$ can be either a gluon, a quark 
or an anti-quark in our partonic Monte Carlo.  The SM amplitudes were 
generated using Madgraph~\cite{mad} in the framework of Helas~\cite{helas} 
routines. The NCQED interactions arising from the Lagrangian
(\ref{action}) were implemented as subroutines and were included
accordingly. We consistently took into account the effect of all
interferences between the NCQED and the SM amplitudes, and considered a 
center--of--mass energy of 14 TeV and an integrated luminosity of 
100 fb$^{-1}$ for LHC.

The process (\ref{jj}) receives contributions from the NCQED 
$\gamma \gamma \gamma$ and $\gamma\gamma\gamma\gamma$ vertices. 
In order to reduce the enormous QCD background we must exploit the
characteristics of the WBF reactions.  The main feature of WBF
processes is a pair of very far forward/backward tagging jets with
significant transverse momentum and large invariant mass between them.
Therefore, we required that the jets should comply with
\footnote{
Another advantage of the choice of cuts (\ref{cuts_jj1}) is the following: 
if we assume that possible non--commutative quark-quark-photon interactions 
have an exponential dependence involving the real antisymmetric 
matrix $C_{\mu \nu}$, like the NCQED lepton-lepton-photon interaction 
given by $\gamma_\mu f(p_1) \bar{f}(p2) \; : \; 
i g \gamma^\mu \exp(\frac{ip_1Cp_2}{2 \Lambda^2})$,
then the effects of these non--commutative quark-quark-photon interactions
are negligible because the set of cuts (\ref{cuts_jj1}) makes
$\exp(\frac{ip_{\rm quark}Cp_{\rm quark}}{2 \Lambda^2}) \to 1 \;,$
allowing us to consider only SM quark-quark-photon interactions in 
our analysis.}

\begin{eqnarray}
&&p_{T}^{j_{1(2)}} >  40~ (20)  \; \text{GeV} \;\;\;\; \hbox{,} \;\;\;\;
|\eta_{j_{(1,2)}}| <  5.0  \; ,
\nonumber \\
&& |\eta_{j_{1}} - \eta_{j_{2}}| >  4.4  \;\;\;\; \hbox{,} \;\;\;\;\;\;
\eta_{j_{1}} \cdot \eta_{j_{2}} < 0  \;\;\;\; \hbox{and} 
\label{cuts_jj1} \\
&& \Delta R_{jj} >  0.7\; .
 \nonumber 
\end{eqnarray}

Furthermore, the photons are central, typically being between the
tagging jets. So, we require that the photons satisfy
\begin{eqnarray}
&& E_{T}^{\gamma_{(1,2)}}  >   25  \; \text{GeV} 
 \;\;\;\; \hbox{,} \;\;\;\;
|\eta_{\gamma_{(1,2)}}|   <  2.5  \; ,
\nonumber 
\label{cuts_jj2}
\\
&& \text{min}\{\eta_{j_{1}}, \eta_{j_{2}} \} + 0.7 <
 \eta_{\gamma_{(1,2)}} 
< \text{max}\{ \eta_{j_{1}}, \eta_{j_{2}} \} - 0.7 \; , 
 \\
&& \Delta R_{j\gamma} > 0.7  \;\;\;\; \hbox{and}
\;\;\;\; \Delta R_{\gamma \gamma} > 0.4
\; .
 \nonumber 
\end{eqnarray}

\begin{figure}
\protect
\centerline{\mbox{\psfig{file=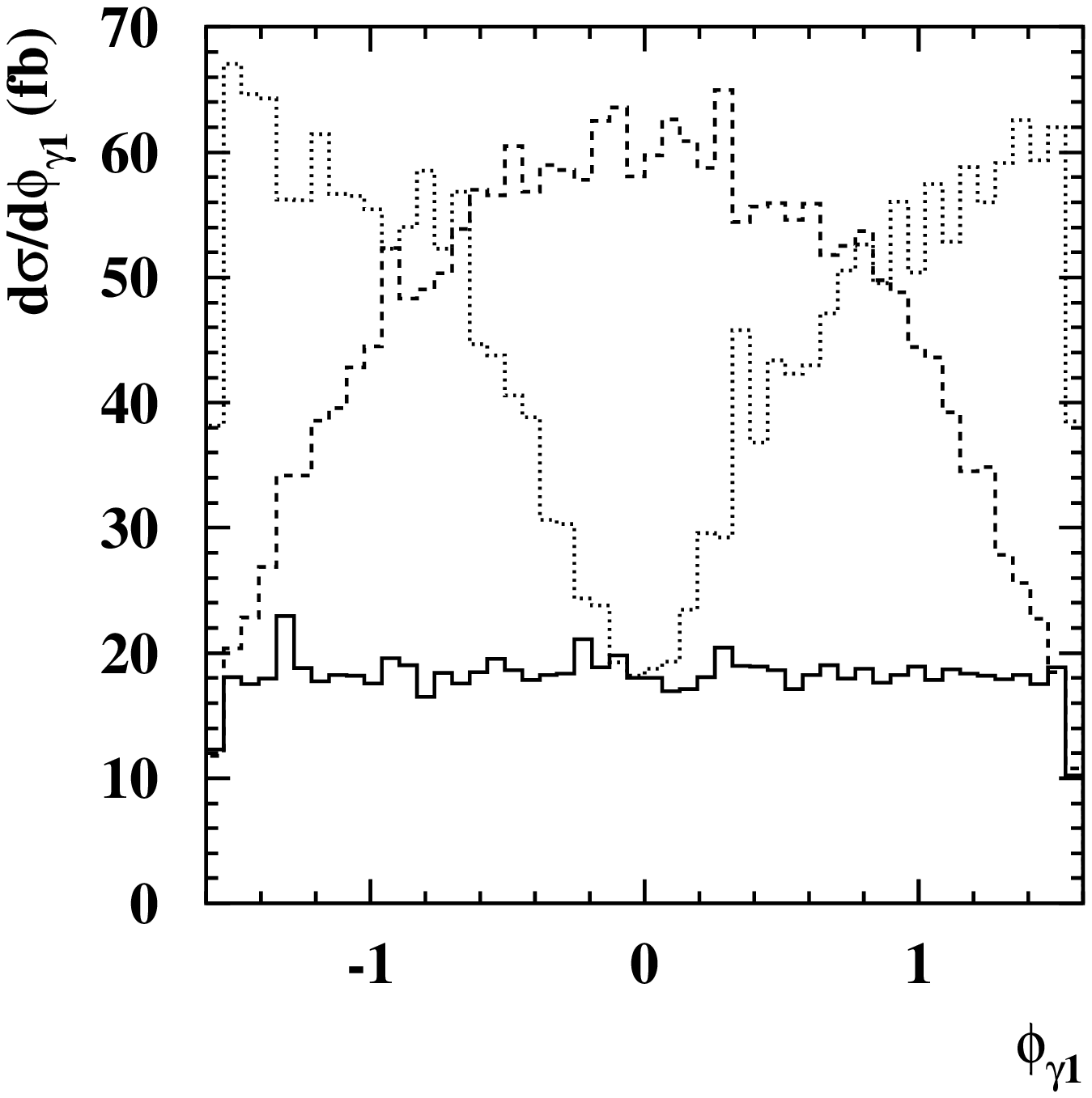,angle=0,width=0.5\textwidth}}
            \mbox{\psfig{file=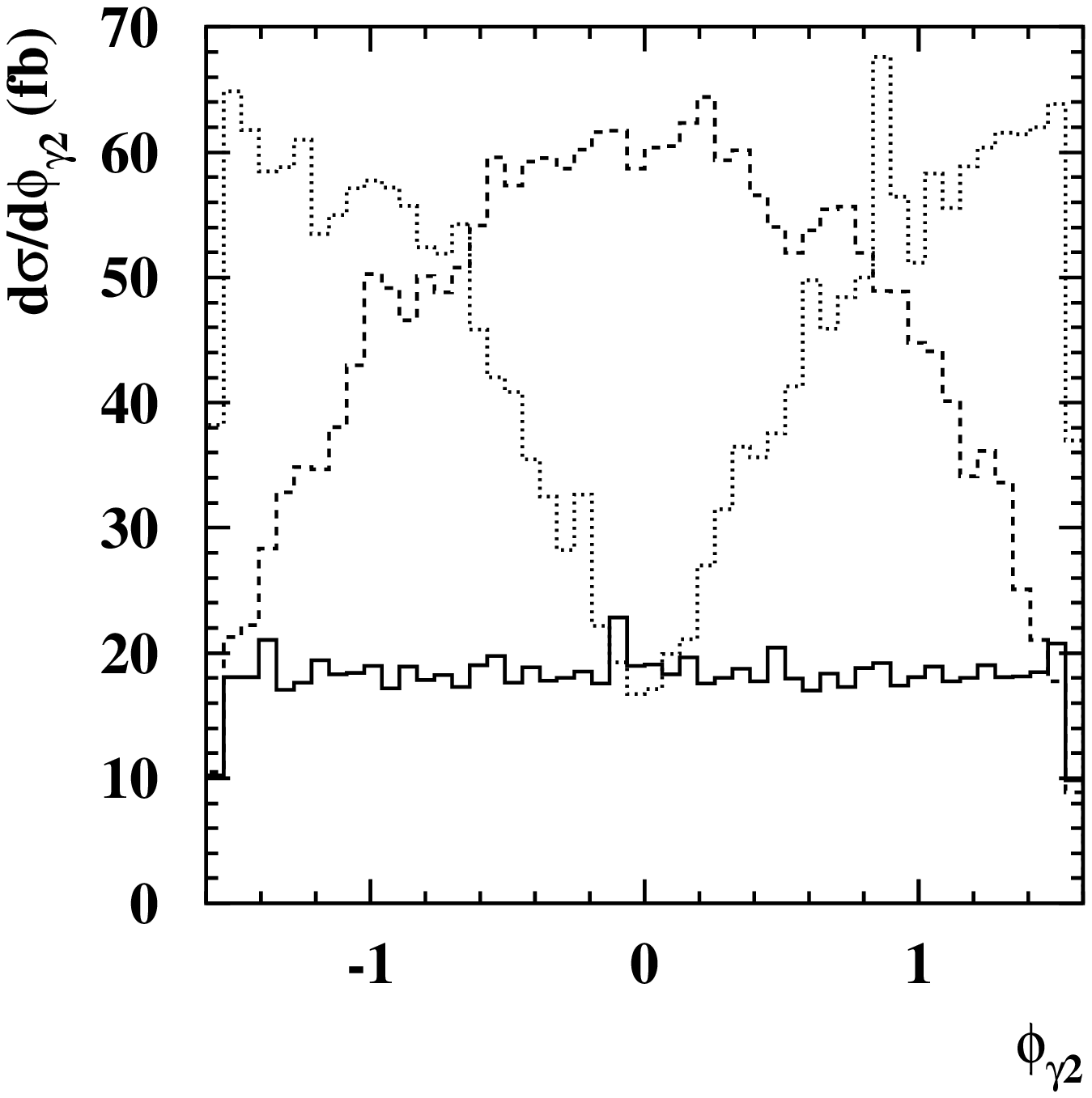,angle=0,width=0.5\textwidth}}}
\centerline{\mbox{\psfig{file=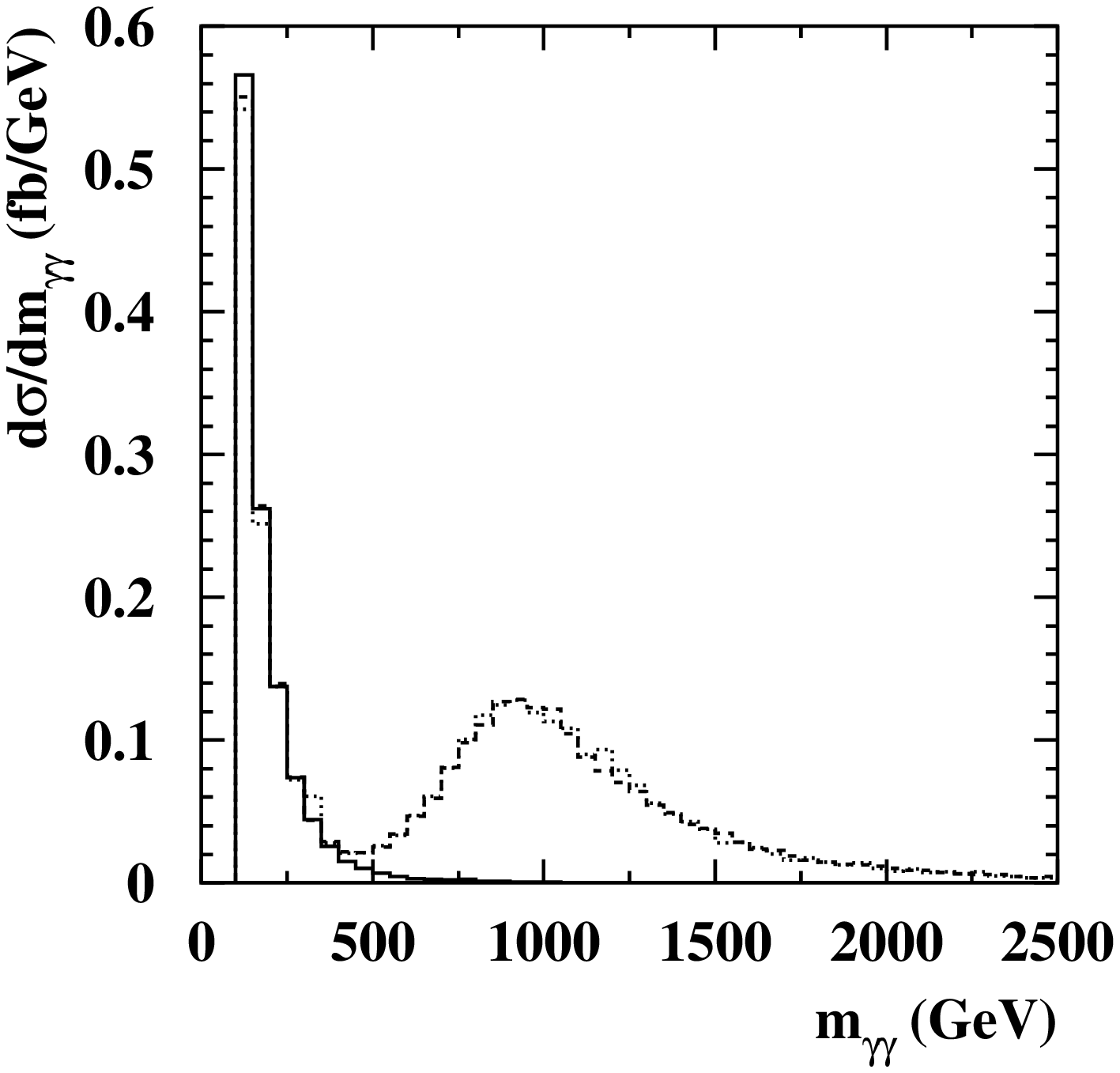,angle=0,width=0.5\textwidth}}}
\vskip -1. cm 
\caption{Azimuthal angle distribution of the most energetic final photon
($\Phi_{\gamma1}$), azimuthal angle distribution of the least energetic 
final photon ($\Phi_{\gamma2}$), and invariant mass distribution of the 
$\gamma\gamma$ pair ($m_{\gamma\gamma}$)
for the reaction $p  p \to \gamma \gamma j j$. The full line
represents the SM distribution. The NCQED contribution was obtained
for $\Lambda = 250$ GeV, and the angles $\alpha, \beta$ and $\gamma$
equal to $\pi/2$. The dashed (dotted) line represents the 
space-space (space-time) non--commutativity case, discussed in the text.}
\label{fig02}
\end{figure}

Several kinematic distributions were evaluated in order to reduce the SM 
background with minimum impact over the NCQED signal. Better results
were observed in three distributions: 
the azimuthal angle distribution of the most energetic final photon
($\Phi_{\gamma1}$), the azimuthal angle distribution of the least 
energetic final photon ($\Phi_{\gamma2}$), and
the invariant mass distribution of the $\gamma\gamma$ pair
($m_{\gamma\gamma}$), presented in Fig.~\ref{fig02}. 
It is interesting to notice that the presence of an NCQED signal 
changes the behavior of the azimuthal angle distribution of a final
photon, while other known examples of new physics can not produce 
similar effect.
However, an impressive reduction of the SM background with small effect
over the NCQED signal can be achieved by 
a cut in the invariant mass distribution of the $\gamma\gamma$ pairs. 
As illustrated in Fig.~\ref{fig02}, the invariant mass
distribution for the SM background contribution is a decreasing
function of the $\gamma\gamma$ invariant mass while the NCQED
contribution first increases with the $\gamma\gamma$ invariant mass
reaching its maximum value at $m_{\gamma \gamma} \sim 850$ GeV and
then decreases. Consequently, in order to enhance the WBF signal for
the NCQED $\gamma\gamma\gamma$ and $\gamma\gamma\gamma\gamma$ 
couplings we imposed the following additional cut in the
diphoton invariant mass spectrum
\begin{eqnarray}
400 \text{ GeV } \leq m_{\gamma \gamma} \leq  2500
\text{ GeV.}
\label{cuts_jj3}
\end{eqnarray}
The results presented in Fig. \ref{fig02} were obtained using
$\sqrt{\hat{s}}$ as the factorization scale in the parton distribution
functions, and the renormalization scale ($\mu_R$) used in the evaluation of
the QCD coupling constant [$\alpha_S(\mu_R)$] was defined such that 
$\alpha_s^2(\mu_R)= \alpha_s(p_T^{j1})\alpha_s(p_T^{j2})$, where 
$p_T^{j1}$ and $p_T^{j2}$ are the transverse momentum of the tagging jets.  

The evaluation of the SM background ($\sigma_{\text{sm}}$) deserves
some special care since it has a large contribution from QCD
subprocesses whose size depends on the choice of the renormalization
scale used in the evaluation of the QCD coupling constant,
$\alpha_s(\mu_R)$, as well as on the factorization scale $\mu_F$ used
for the parton distribution functions.  To estimate the uncertainty
associated with these choices, we have reproduced the procedure used 
in Ref.\ \cite{Eboli:2003nq}, computing $\sigma_{\text{sm}}$
for two sets of renormalization scales, which we label as
$\mu_{R1,2}(\xi)$, and for several values of $\mu_F$.  $\mu_{R1}(\xi)$
is defined such that $\alpha_s^2(\mu_{R1}(\xi))= \alpha_s(\xi
p_T^{j1})\alpha_s(\xi p_T^{j2})$ where $p_T^{j1}$ and $p_T^{j2}$ are
the transverse momentum of the tagging jets and $\xi$ is a free
parameter varied between 0.1 and 10. The second choice of
renormalization scale set is $\mu_{R2} (\xi) =\xi \sqrt{\hat s}/2$,
with $\sqrt{\hat{s}}$ being the subprocess center--of-mass energy.

\begin{table}
\begin{tabular}{||c||c|c|c||c|c|c||}
\hline 
\hline 
&  \multicolumn{6}{c||}{$\sigma_{\text{sm}}$ (fb)} \\ 
\hline
& \multicolumn{3}{c||}{$\mu_R=\mu_{R1}(\xi)$ } 
& \multicolumn{3}{c||}{$\mu_R=\mu_{R2}(\xi)$ } \\ 
\hline
$\xi$ & 
$\mu_F=\sqrt{\hat s} $& $\mu_F=p^T_{\rm min}$ 
& $\mu_F=\sqrt{\hat s}/10 $ & 
$\mu_F=\sqrt{\hat s} $& $\mu_F=p^T_{\rm min}$ 
& $\mu_F=\sqrt{\hat s}/10 $\\
\hline
\hline 
0.10 & 3.2 & 5.3 & 4.1 & 1.3 & 2.2 & 1.7\\
\hline 
0.25 & 2.2 & 3.6 & 2.8 & 1.1 & 1.9 & 1.4\\ 
\hline 
1.00 & 1.4 & 2.4 & 1.9 & 0.91 & 1.5 & 1.2\\ 
\hline 
4.00 & 1.1 & 1.8 & 1.4 & 0.78 & 1.3 & 1.0\\ 
\hline 
10.0 & 0.94 & 1.6 & 1.2 & 0.71 & 1.2 & 0.96
\\ 
\hline 
\hline 
\end{tabular}
\medskip
\caption{Results for $\sigma_{\text{sm}}$ for process 
Eq.~(\protect{\ref{jj}}); see text for details.
All results include the effect of the cuts in 
Eq.~(\ref{cuts_jj1}), (\ref{cuts_jj2}) and (\ref{cuts_jj3}) as well as
photon detection and jet-tagging efficiencies.}
\label{tabsm_aa}
\end{table}

For now on our results will be presented  assuming a 85\% detection 
efficiency of isolated photons and jet-tagging.  With this the efficiency 
for reconstructing the final state $j + j + \gamma + \gamma$ is 
$(0.85)^4 \approx$ 52\% which is included in our results .
In Table \ref{tabsm_aa} we list $\sigma_{\text{sm}}$ for the two sets
of renormalization scales 
and for three values of the factorization scale $\mu_F=\sqrt{\hat s}$,
$\sqrt{\hat s}/10$, and $p^T_{\rm min}$ where $p^T_{\rm min}={\rm min}
(p_T^{j1},p_T^{j2})$. 
As shown in this table, we find that the
predicted SM background can change by a factor of $\sim 8$ depending
on the choice of the QCD scales.  These results indicate that to
obtain meaningful information about the presence of NCQED
$\gamma \gamma \gamma$ and $\gamma \gamma \gamma \gamma$
couplings one cannot rely on the theoretical evaluation of the
background.  Instead one should attempt to extract the value of the SM
background from data in a region of phase space where no signal is
expected and then extrapolate to the signal region.

In looking for the optimum region of phase space to perform this
extrapolation, one must search for kinematic distributions for which
(i) the shape of the distribution is as independent as possible of the
choice of QCD parameters. Furthermore, since the electroweak and QCD
contributions to the SM backgrounds are of the same order,
\footnote{The electroweak contribution to the total SM background
is approximately 25\% for $\mu_{R1}(\xi=1)$ and $\mu_F = \sqrt{\hat{s}}$.}
this requires that (ii) the shape of both electroweak and QCD
contributions are similar. Several kinematic distributions verify
condition (i), for example, the azimuthal angle separation of the two
tagging jets which was proposed in Ref.~\cite{Eboli:2000ze} to reduce
the perturbative QCD uncertainties of the SM background estimation for
invisible Higgs searches at LHC.  However, the totally different shape
of the electroweak background in the present case, renders this
distribution useless.

\begin{figure}
\protect
\centerline{\mbox{\psfig{file=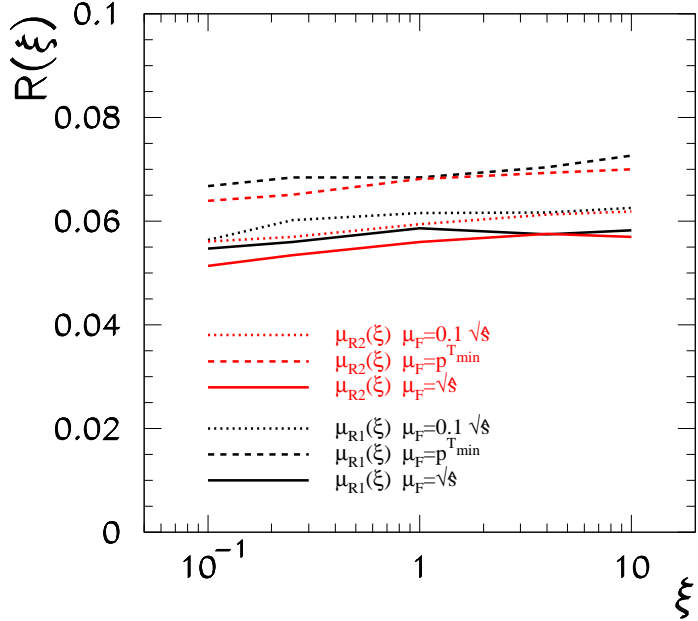,angle=0,width=0.6\textwidth}}}
\caption{Ratio $R(\xi)$ defined in Eq.~(\ref{rxijj})  
for the process $p p \to \gamma  \gamma  j  j$ at LHC.}
\label{fig03}
\end{figure}

We found that the best sensitivity is obtained by using the
$\gamma\gamma$ invariant mass.  As can be seen in
Fig.~\ref{fig02}, the shape of the SM distribution is quite
independent of the choice of the QCD parameters.  As a consequence
most of the QCD uncertainties cancel out in the ratio
\begin{equation}
R(\xi)=\frac{\sigma({400\text{ GeV }<m_{\gamma \gamma}<2500\text{ GeV }}) }
{\sigma({100\text{ GeV }<m_{\gamma \gamma}<400\text{ GeV }}) }\;\;.
\label{rxijj}
\end{equation}
This fact is illustrated in Fig.~\ref{fig03} where we plot the value
of the ratio $R(\xi)$ for different values of the renormalization and
factorization scales.  The ratio $R$ is almost invariant under changes
of the renormalization scale, showing a maximum variation of the order
of $\pm6$\% for a fixed value of the factorization scale. On the other
hand, the uncertainty on the factorization scale leads to a maximum
variation of 12\% in the background estimation.  We have also verified
that different choices for the structure functions do not affect these
results.

Thus the strategy here proposed is simple: the experiments should
measure the number of events in the $\gamma\gamma$ invariant mass
window $100 < m_{\gamma\gamma}< 400$ GeV and extrapolate the results
for higher invariant masses using perturbative QCD.  According to the
results described above we can conservatively assign a maximum ``QCD''
uncertainty ($ {\rm QCD}_{\rm unc}$) of $\pm$ 15\% to this
extrapolation.

In order to estimate the attainable sensitivity to NCQED
we assume that the observed number of events is compatible
with the expectations for $\mu_{R1}(\xi=1)$ and $\mu_F=\sqrt{\hat s}$,
so the observed number of events in the signal region coincides with
the estimated number of background events obtained from the
extrapolation of the observed number of events in the region where no
signal is expected; for this choice the number of expected background
events is $N_{\rm sm}=\sigma_{\text{sm}} {\cal L}$ where ${\cal L}$
stands for the integrated luminosity.  For an integrated luminosity of
100 fb$^{-1}$ for LHC, this corresponds to $N_{\rm back}=143$.
Moreover, we have added in quadrature the statistical error and the
QCD uncertainty associated with the backgrounds. Therefore, the 95\% C.L.
limits on $\Lambda$ can be obtained from the condition
\begin{equation}
N_{\rm NCQED}\,=\, {\cal L} \,\times \, \sigma_{\rm NCQED} 
\leq  1.95 \sqrt{N_{\rm sm}+ (N_{\rm sm}\times {\rm QCD}_{\rm
    unc})^2} \; ,
\label{condition1}
\end{equation}
where $N_{\rm NCQED}$ ($\sigma_{\rm NCQED}$) is the maximum number of events 
(cross section) deviation due to the NCQED contribution, so 
$N_{\rm observed} = N_{\rm sm} \pm N_{\rm NCQED}$. Once we have 
${\cal L} = 100 {\rm fb}^{-1}$ and $N_{\rm back}=143$, equation 
(\ref{condition1}) turns out to be
\begin{equation}
\sigma_{\rm NCQED} ({\rm fb}) \leq  
0.0195 \times \sqrt{143+ (143 \times {\rm QCD}_{\rm unc})^2} \; .
\label{condition2}
\end{equation}

For the sake of completeness we show the results on the expected
sensitivity using purely statistical errors and for two values of
${\rm QCD}_{\rm unc}$: our most conservative estimate [15 \%], and a
possible reduced uncertainty (7.5 \%), which could be attainable provided
NLO QCD calculations are available. Therefore the NCQED deviation
should not be greater than the values presented in Table \ref{tab_NCQED}.

\begin{table}
\begin{tabular}{||c||c||c||}
\hline 
\hline 
${\rm QCD}_{\rm unc}$ & $\sigma_{\rm NCQED} ({\rm fb})$ & $N_{\rm NCQED}$ \\ 
\hline
0 & 0.23 & 23 \\ 
\hline
7.5\% & 0.31 & 31 \\ 
\hline
15\% & 0.48 & 48 \\ 
\hline 
\hline 
\end{tabular}
\medskip
\caption{95\% C.L. maximum cross section and number of events 
deviation due to the NCQED contribution.}
\label{tab_NCQED}
\end{table}

Once we have fixed $\phi$ by settling $\beta =\pi/2$ 
in equation (\ref{matrixC}), the antisymmetric matrix 
get parametrized by two angles: the angle $\alpha$ related 
to the space-time non--commutativity, and the angle $\gamma$  
related to the space-space non--commutativity.
Therefore, in order to perform our analysis we consider two cases:
\begin{itemize}
\item (i) the space-space non--commutativity, where the elements $C_{0i}$
($i=1,2,3$) in equation (\ref{matrixC}) are assumed to be $0$, 
and the angle $\gamma$ are assumed to be either $0$, $\pi/4$ or $\pi/2$;
\item (ii) the space-time non--commutativity, where the elements $C_{ij}$
($i,j=1,2,3$) in equation (\ref{matrixC}) are assumed to be $0$, 
and the angle $\alpha$ are assumed to be  either $0$, $\pi/4$ or $\pi/2$.
\end{itemize}
\begin{figure}
\protect
\centerline{\mbox{\psfig{file=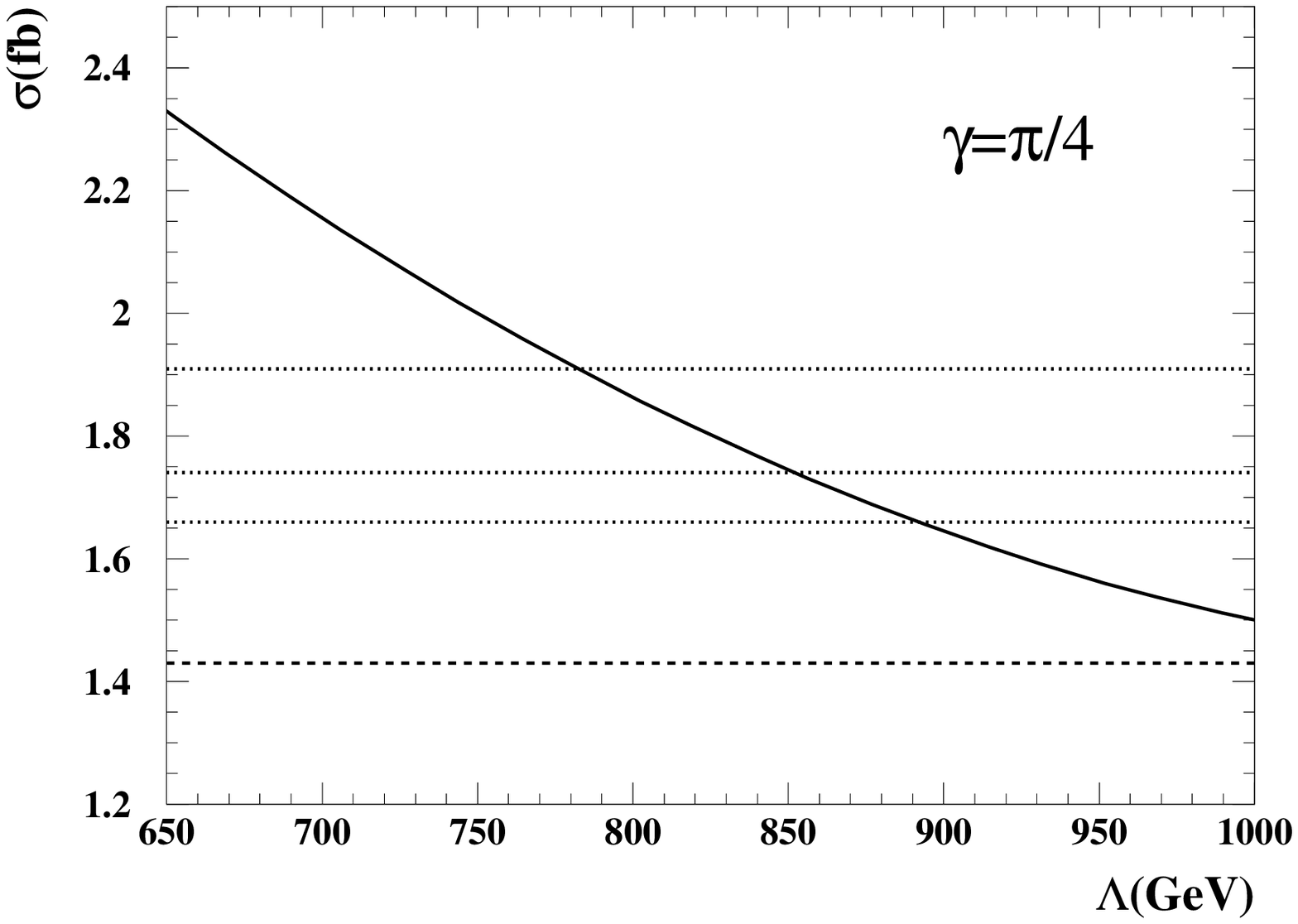,angle=0,width=0.55 \textwidth ,clip=}}
\mbox{\psfig{file=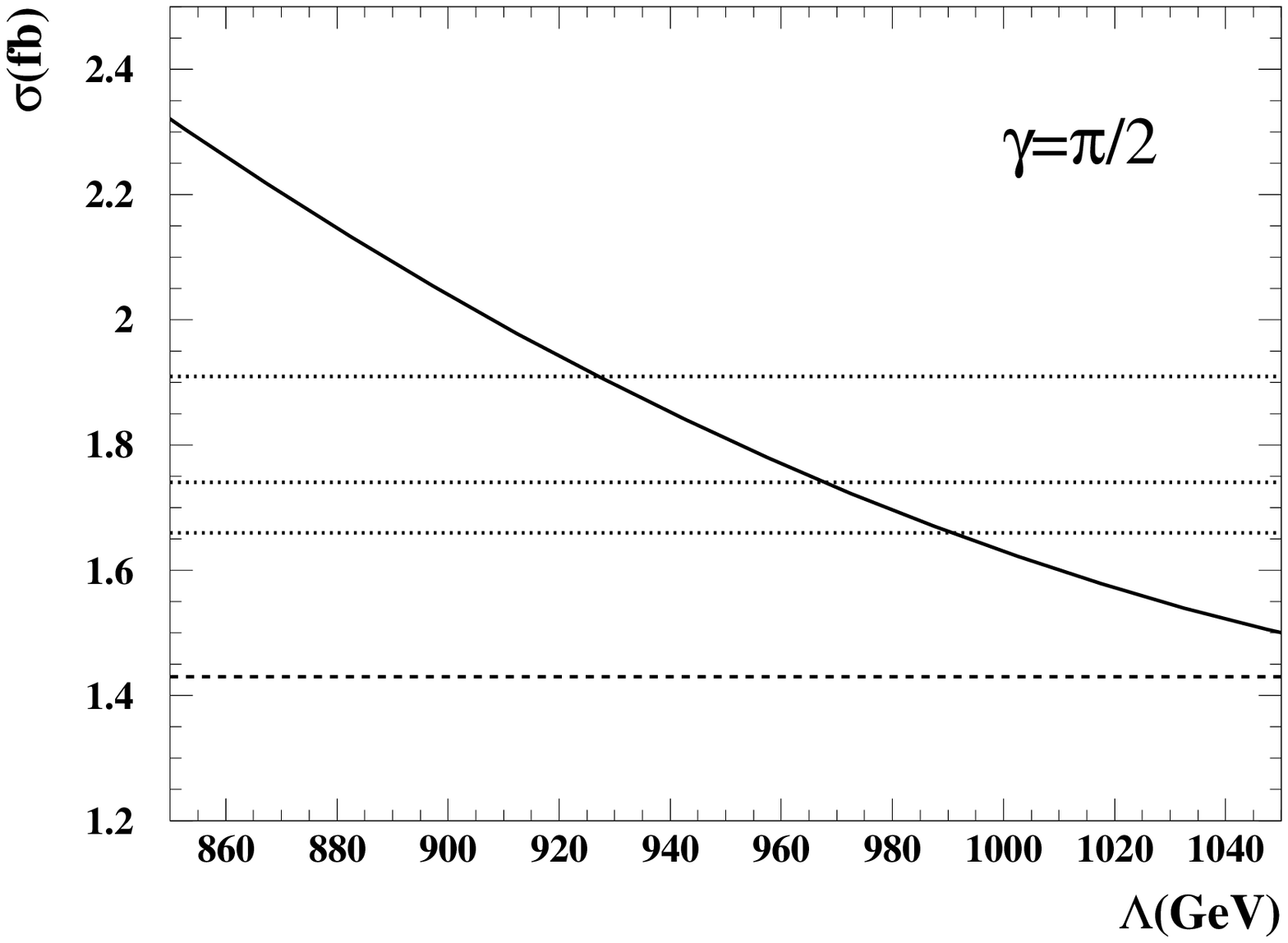,angle=0,width=0.55 \textwidth ,clip=}}}
\vskip -1. cm 
\caption{Cross section results including the effect of the cuts in 
Eq.~(\ref{cuts_jj1}), (\ref{cuts_jj2}) and (\ref{cuts_jj3}) as well as
photon detection and jet-tagging efficiencies. 
The solid line is the cross section including space-space NCQED effects,
the dashed line is the SM cross section, and the upper (medium) [lower] 
dotted line represents the 95\% C.L. upper limit for an
QCD uncertainty of 15\% (7.5\%) [0].}
\label{fig04}
\end{figure}

In order to determine which case of NCQED is being observed, one can
use the azimuthal angle distribution of a final photon. As illustrated 
in Fig.~\ref{fig02},  the azimuthal angle distribution of either the most
or the least energetic final photon for the SM background contribution 
is a flat function in the range $|\Phi_\gamma| < \frac{\pi}{2}$, while for the 
space-time (space-space) non--commutativity signal contribution, the
distribution is increased for $|\Phi_\gamma| \to \frac{\pi}{2}(0)$.
\footnote{We have checked that this angular behavior is preserved 
for  $\Lambda \simeq 1$ TeV after the inclusion of the cut 
(\ref{cuts_jj3}) in our evaluations.} 

The evaluation of the cross section of the reaction (\ref{jj}), including the 
effect of the cuts in Eq.~(\ref{cuts_jj1}), (\ref{cuts_jj2}) and 
(\ref{cuts_jj3}) as well as photon detection and jet-tagging efficiencies,
is now done including the effect of the diagrams in
Fig.~(\ref{fig01}), for the cases (i) and (ii) described above.

The results for the space-space non--commutativity, case (i), are
presented in Fig.~(\ref{fig04}), for $\gamma=\pi/4$, and $\pi/2$. 
No limits on $\Lambda$ could be obtained for $\gamma=0$.
Our analysis shows a better sensitivity for the angle $\gamma = \pi/2$, 
allowing us to impose a lower limit on the NCQED scale of 
$\Lambda \gtrsim$ 990 GeV if the QCD uncertainty discussed above is not 
considered. The limit changes to $\Lambda\gtrsim$  930 (960) GeV
for a  QCD uncertainty of 15\% (7.5\%). For $\gamma = \pi/4$, the lower 
limit on the NCQED scale is  $\Lambda\gtrsim$ 780 (850) [900] GeV for a 
QCD uncertainty of 15\% (7.5\%) [0\%].

\begin{figure}
\protect
\centerline{\mbox{\psfig{file=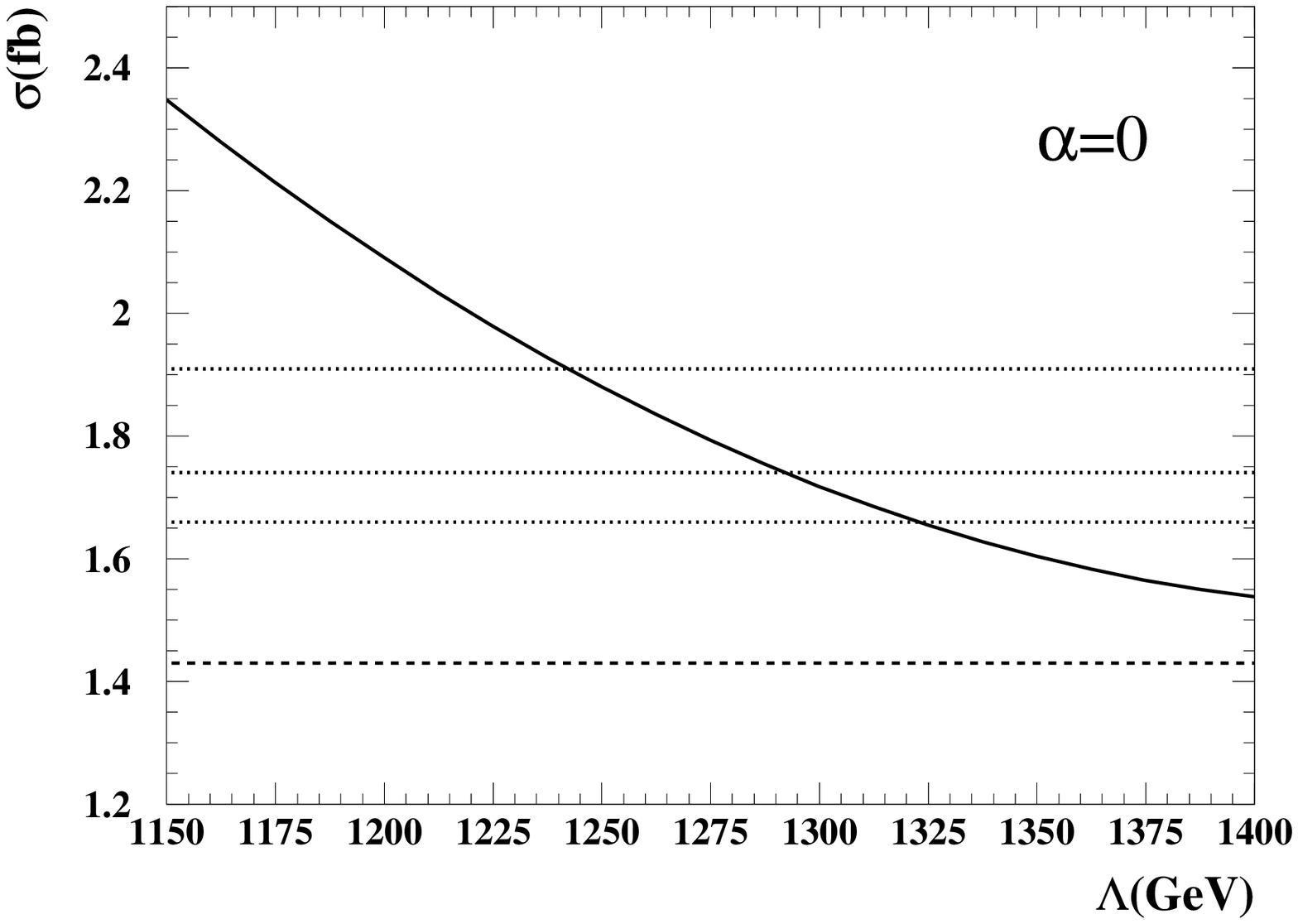,angle=0,width=0.55\textwidth,clip=}}}
\centerline{\mbox{\psfig{file=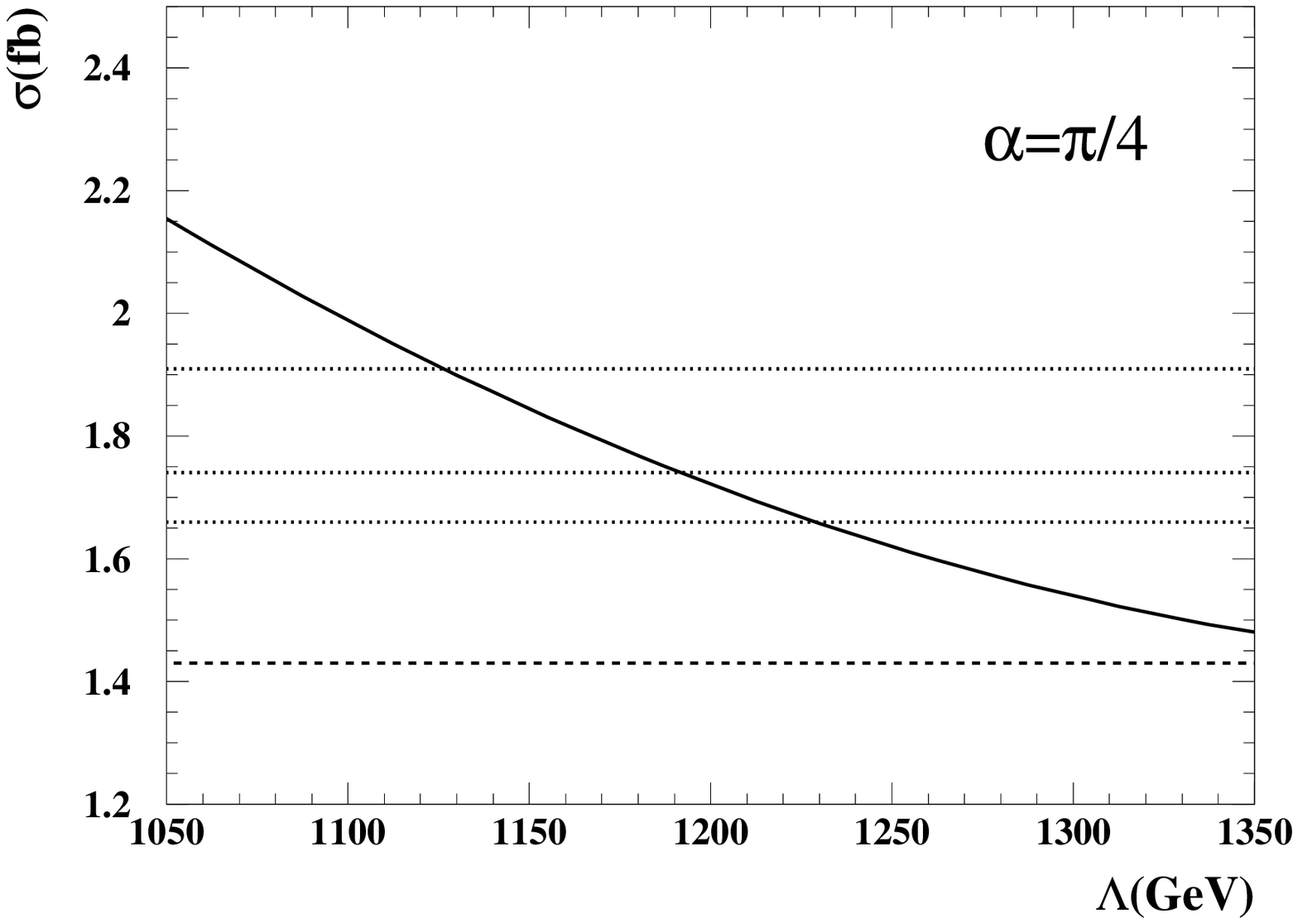,angle=0,width=0.55\textwidth,clip=}}
            \mbox{\psfig{file=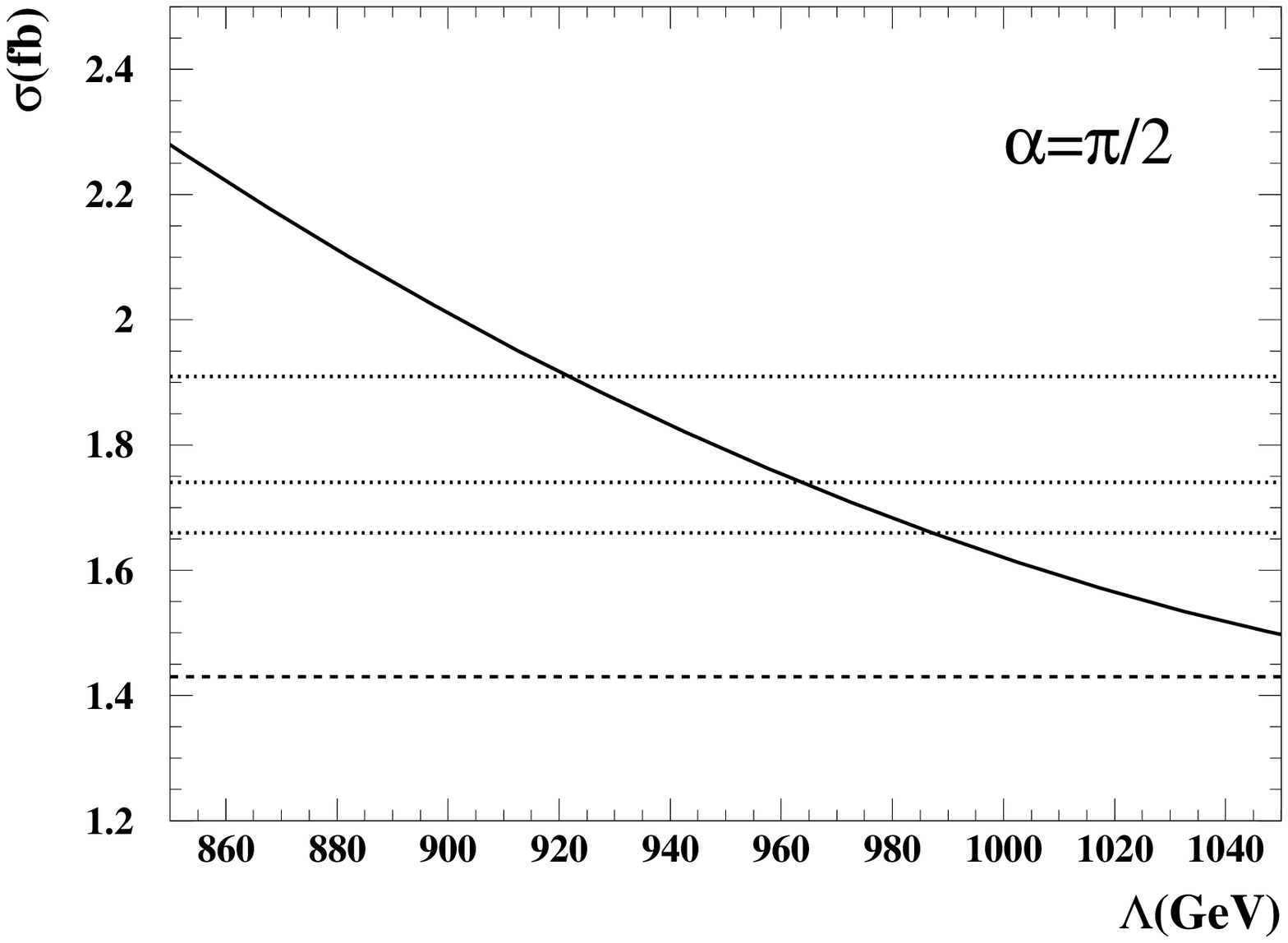,angle=0,width=0.55\textwidth,clip=}}}
\vskip -1. cm 
\caption{Cross section results including the effect of the cuts in 
Eq.~(\ref{cuts_jj1}), (\ref{cuts_jj2}) and (\ref{cuts_jj3}) as well as
photon detection and jet-tagging efficiencies. 
The solid line is the cross section including space-time NCQED effects,
the dashed line is the SM cross section, and the upper (medium) [lower] 
dotted line represents the 95\% C.L. upper limit for an
QCD uncertainty of 15\% (7.5\%) [0].}
\label{fig05}
\end{figure}

On the other hand, the results for the space-time non--commutativity, 
case (ii), are presented in Fig.~(\ref{fig05}), for
$\alpha=0$, $\pi/4$, and $\pi/2$, respectively.
Our analysis shows a better sensitivity for the angle $\alpha = 0$, 
allowing us to impose a lower limit on the NCQED scale of 
$\Lambda \gtrsim$ 1320 GeV if the QCD uncertainty discussed above is not 
considered. The limit changes to $\Lambda\gtrsim$  1290 (1230) GeV
for a  QCD uncertainty of 15\% (7.5\%). For $\alpha = \pi/4$, the lower 
limit on the NCQED scale is  $\Lambda\gtrsim$ 1130 (1190) [1230] GeV for a 
QCD uncertainty of 15\% (7.5\%) [0\%] and for $\alpha = 0$, the lower 
limit on the NCQED scale is  $\Lambda\gtrsim$ 920 (960) [990] GeV for a 
QCD uncertainty of 15\% (7.5\%) [0\%].

\section{Conclusions}
\label{sec4}

In this work we investigated the potential for LHC to probe the 
photonic 3- and 4- point functions that appears in  NCQED through the
analysis of the process (\ref{jj}).
Even though we assumed that the quark-quark-photon interactions 
make part of  the SM background due to our choice of kinematical cuts, 
this process is sensitive  for space-space as well as for space-time 
non--commutativity.  Our main results are presented in 
Fig.~\ref{fig04} and Fig.~\ref{fig05} where the space-space and 
space-time NCQED effects are observed. 

For the space-space non--commutativity, our study shows a better sensitivity
for the angle $\gamma = \pi/2$, allowing us to impose a lower limit on
the NCQED scale $\Lambda$ in the range 930 GeV $\leq \Lambda \leq$ 990 GeV,
depending on the perturbative QCD uncertainties considered.
Regarding the space-time non--commutativity, the process is more sensitive for 
the angle $\alpha =0$,  where a lower limit on
the NCQED scale $\Lambda$ in the range 1230 GeV $\leq \Lambda \leq$ 1320 GeV,
depending on the perturbative QCD uncertainties considered, could be imposed.

Therefore, this work shows that LHC may be a good place to test NCQED
via the study of the  process (\ref{jj}). We have shown that LHC is able 
to probe both space-space and space-time non--commutativity. A better
sensitivity is expected for the  space-time non--commutativity, where
the NCQED scale $\Lambda$ can be tested up to $\Lambda \sim 1.25$ TeV.

%%%%%%%%%%%%%%%%%%%%%%%%%%%%%%%%%%%%%%%%%%%%%%%%%%%%%%%%%%%%%%%%%%%%%%

\acknowledgments
The authors would like to thank R. F. Ribeiro for useful discussions. 
S.~M.~Lietti thanks  the hospitality received
at UFPb during the early stage of this work.
This work was supported by Conselho Nacional de Desenvolvimento
Cient\'{\i}fico e Tecnol\'ogico (CNPq), and by Funda\c{c}\~ao de Amparo
\`a Pesquisa do Estado de S\~ao Paulo (FAPESP).

%--- References ---
\def\MPL #1 #2 #3 {Mod. Phys. Lett. A {\bf#1},\ #2 (#3)}
\def\NPB #1 #2 #3 {Nucl. Phys. {\bf#1},\ #2 (#3)}
\def\PLB #1 #2 #3 {Phys. Lett. B {\bf#1},\ #2 (#3)}
\def\PR #1 #2 #3 {Phys. Rep. {\bf#1},\ #2 (#3)}
\def\PRD #1 #2 #3 {Phys. Rev. D {\bf#1},\ #2 (#3)}
\def\PRL #1 #2 #3 {Phys. Rev. Lett. {\bf#1},\ #2 (#3)}
\def\RMP #1 #2 #3 {Rev. Mod. Phys. {\bf#1},\ #2 (#3)}
\def\NIM #1 #2 #3 {Nuc. Inst. Meth. {\bf#1},\ #2 (#3)}
\def\ZPC #1 #2 #3 {Z. Phys. {\bf#1},\ #2 (#3)}
\def\EJPC #1 #2 #3 {Eur. Phys. J. C {\bf#1},\ #2 (#3)}
\def\IJMP #1 #2 #3 {Int. J. Mod. Phys. A {\bf#1},\ #2 (#3)}
\def\JHEP #1 #2 #3 {J. High Energy Phys. {\bf#1},\ #2 (#3)}

\end{document}